
:
\magnification \magstep1
\centerline  {\bf New Features of the Mandelstam-Leibbrandt Lightcone Gauge}
\vskip 2cm
\centerline { A. Andra\v si }
\centerline  {\it `Rudjer Bo\v skovi\' c' Institute, Zagreb, Croatia }
\vskip 4cm

\beginsection Abstract

 This is about new unexpected features of the Mandelstam-Leibbrandt
 prescription found as applied to spacelike Wilson lines. The regularization
 parameter $  \omega $ in the M-L denominator for the spurious poles has
 to be kept throughout the calculation till the very end or else the
 integrals do not make sense. We get various `ambiguous' terms of the
 form $ {\omega}^{-{\epsilon\over2}}{\epsilon}^{-2} $ which are not
 controlled by any sort of Ward identity. These terms cancel out in
 the sum and the final result is independent of $ \omega $. However, for
 the selfenergy on the spacelike Wilson line we obtain an unexpected
 double pole at $ \epsilon=0 $, using dimensional regularization in
 $ 4-\epsilon $ dimensions.

 \vskip 5cm

Electronic address:(internet)andrasi@thphys.irb.hr
\vfill \eject

\beginsection   1. Introduction

Non-covariant gauges found applications in a wide range of fields,
such as QCD,   supersymmetric Yang-Mills and string theories. A
favourite among them is Mandelstam's lightcone gauge [1] defined
by two lightlike vectors satisfying
$$ n^2=n^{*2}=0,\; \; n\cdot n^*=2,  \eqno(1)  $$ and the propagator
$$ {{\delta_{ab}}\over{k^2+i\eta}}\{g_{\mu\nu}-{{k_{\mu}n_{\nu}+
k_{\nu}n_{\mu}}\over{n\cdot k+i\omega n^*\cdot k}}\}. \eqno(2) $$
However, this prescription leads to non-local counterterms [2],
does not satisfy the optical theorem [3] and in order to satisfy
the Piguet-Sibold equation [4] requires a more precise definition
[5]. A convenient laboratory to study the prescription is the
Wilson loop [6]. Here we take the triangle Wilson loop with two
sides in the direction of the two lightlike vectors and the
base in spacelike direction, defined by
$$ x^1_{\mu}=n_{\mu}^*L -n_{\mu}Lt  $$
$$ x^2_{\mu}=v_{\mu}2L(1-s)   $$
$$ x^3_{\mu}=n_{\mu}^*Lt  \eqno(3) $$
where
$$ n^* _{\mu}=(1,0,0,1)  $$
$$ n_{\mu}=(1,0,0,-1)  $$
$$ v_{\mu}=(0,0,0,1)  $$
$$ 0 \leq t \leq 1 $$
$$ 0 \leq s \leq 1.\eqno(4)   $$

\beginsection 2. Self Energy Diagram

There are only two diagrams contributing to order $ g^2 $
in the lightcone gauge, the vertex graph where the gluon
propagates between the base in the direction $ v_{\mu} $
and the side in the direction $ n^*_{\mu} $, and the self energy
graph on the base in the direction $ v_{\mu} $.
The selfenergy diagram in the momentum space contributes
$$ W_2=ig^2C_R{ \int}\,{{d^nk}\over{{(2\pi)^n}}}{1\over{k^2+i\eta}}
{{k_-}\over{k_+ +i\omega k_-}}{1\over{k_3^2}}(\cos 2Lk_3 -1)
\eqno(5) $$
where
$$ k_-=n^*\cdot k=k_0 -k_3, $$
$$ k_+=n\cdot k=k_0+k_3. \eqno(6) $$
Closeing the contour in the upper $ k_0 $ half plane, we meet
two poles
$$ k_0=-k+i\eta $$
$$ k_0=-k_3 +2i\omega k_3 \theta (k_3), \eqno(7) $$
which give
$$ W_2=\pi g^2 C_R(2\pi)^{-n}{ \int}\,
dk_3d^{2-\epsilon}K {1\over k}{{k+k_3}\over{k-k_3+i\omega (k+k_3)}}
{1\over{k_3^2}}(\cos 2Lk_3-1) $$
$$ -4\pi g^2C_R(2\pi)^{-n}{ \int_{0}^{\infty}}\,dk_3{ \int}\,
d^{2-\epsilon}K {1\over{K^2+4i\omega k_3^2-i\eta}}
{1\over{k_3}}(\cos 2Lk_3-1).  \eqno(8) $$
Naively, one would let $ \omega \rightarrow 0 $ before
$ \epsilon =4-n $, in the integrand. Then, the latter integral in
eq.(8) would vanish as a tadpole in the perpendicular
momentum $ K $. However, after the introduction of polar
coordinates  $$ k_3=k\cos \theta =kx, $$
$$ d^{3-\epsilon}k=k^{2-\epsilon}dk(1-x^2)^{-{\epsilon\over2}}
dx  \eqno(9) $$ and integration over $ k $, the first integral
leads to an integral which is not defined for any $ \epsilon $
(we also change the variable of integration $ x^2=y $ ).
Therefore we have to keep $ \omega $ in the integrand and
choose to evaluate the diagram in the strip
$$ 1< \epsilon <2  \eqno(10) $$  in which the same diagram
is regularized by $ \epsilon $ in the Feynman gauge. Then
the self energy graph becomes
$$ W_2=g^2C_R(2\pi)^{1-n}\Gamma (-\epsilon)\cos{{\epsilon
\pi}\over2} (2L)^{\epsilon}{{2{\pi}^{1-{\epsilon\over2}}}\over
{\Gamma (1-{\epsilon\over2})}}\{A+{1\over2}\int_{0}^{1}\,
dy(1-y)^{-{\epsilon\over2}}y^{{{\epsilon -3}\over2}}\} $$
$$ -2g^2C_R(2\pi)^{1-n}E. \eqno(11)  $$
Here two integrals each contain the factor $ \omega ^{-
{\epsilon\over2}}\epsilon^{-2} $  which has no limit as
 $ \omega \rightarrow 0 $  .
$$ A=\int_{0}^{1}\,dy(1-y)^{-{\epsilon\over2}}y^{{{\epsilon-1}
\over2}}[1-y(1-4i\omega)]^{-1} $$ $$ =B({{\epsilon +1}\over2},
1-{\epsilon\over2})\{{{\Gamma({3\over2})\Gamma(-{\epsilon\over2})
}\over{\Gamma({1\over2})\Gamma(1-{\epsilon\over2})}}+
(4\omega e^{{{i\pi}\over2}})^{-{\epsilon\over2}}
{{\Gamma({3\over2})\Gamma({\epsilon\over2})}\over{\Gamma({{1+
\epsilon}\over2})}}\}, \eqno(12) $$
$$ E=\int_{0}^{\infty}\,dk_3 { \int}\,d^{2-\epsilon}K
{1\over{K^2+4i\omega k_3^2-i\eta}}{1\over{k_3}}(\cos 2k_3L
-1)  $$  $$ =\pi ^{1-{\epsilon\over2}}\Gamma ({\epsilon\over2})
\Gamma (-\epsilon)\omega^{-{\epsilon\over2}}e^{-{{i\pi
\epsilon}\over4}} \cos {{\epsilon \pi}\over4} L^{\epsilon}.
\eqno(13)  $$
However these poles cancel in the sum leaving a double pole
in $ \epsilon $ for the self energy graph on the spacelike line.
$$ W_2=({g\over{2\pi}})^2C_R\pi^{{\epsilon\over2}}\Gamma
(-\epsilon)\cos{{\epsilon\pi}\over2 }(2L)^{\epsilon}
{{\Gamma(1+\epsilon)}\over{\Gamma(1+{\epsilon\over2})}}
\{-{2\over{\epsilon}} +{1\over{\epsilon-1}}\} \eqno(14) $$
The same graph in the Feynman gauge contains only single
poles in $ \epsilon $.

\beginsection 3. The Vertex Graph

In order to make the calculations more transparent, we devide
the integrand for the vertex graph into three parts,
$$ W_1=W_1^a+W_1^b+W_1^c \eqno(15)  $$ where
$$ W_1^a=-2ig^2C_R{ \int}\,{{d^nk}\over{(2\pi)^n}}{1\over
{k^2+i\eta}}{1\over{k_3}}{1\over{k_++i\omega k_-}}
(e^{ik_+L}-1) \eqno(16) $$
$$ W_1^b=2ig^2C_R{ \int}\,{{d^nk}\over{(2\pi)^n}}{1\over
{k^2+i\eta}}{1\over{k_3}}{1\over{k_++i\omega k_-}}
(e^{ik_-L}-1) \eqno(17)  $$
$$ W_1^c=2ig^2C_R{ \int}\,{{d^nk}\over{(2\pi)^n}}{1\over
{k^2+i\eta}}{1\over{k_++i\omega k_-}}{1\over {k_3}}
(e^{2iLk_3}-1). \eqno(18) $$
Again we close the contour in the upper $ k_0 $ half plane.
The first part of the integrand does not contain any `ambiguous'
poles in the strip eq.(10), and contributes
$$ W_1^a=g^2C_R(2\pi)^{1-n}\Gamma (-\epsilon)L^{\epsilon}
e^{{{i\pi\epsilon}\over2}}
{{2\pi^{1-{\epsilon\over2}}}\over{\Gamma(1-{\epsilon\over2})}} $$
$$\times \{{2\over{\epsilon}}2^{-{\epsilon\over2}}[1+{{\epsilon^2}\over4}
({{\pi^ 2}\over{12}}-{1\over2}\ln^22)]+\ln2-{5\over{24}}
\epsilon\pi^2 \}. \eqno(19)  $$
The second part of the integrand shows the same features
as the self energy graph. After $ k_0 $ integration it becomes
$$ W_1^b=-g^2C_R(2\pi)^{1-n}\int_{0}^{\infty}\,dk k^{-1
-\epsilon}\int_{-1}^{1}\,dx(1-x^2)^{-{\epsilon\over2}} $$
$$ \times{1\over x}{1\over{1-x+i\omega (1+x)}}(e^{-ik(1+x)L}-1)
{{2\pi^{1-{\epsilon\over2}}}\over{\Gamma(1-{\epsilon\over2})}} $$
$$ +4g^2C_R\pi^{2-{\epsilon\over2}}\Gamma({\epsilon\over2})
\Gamma(-\epsilon)\omega^{-{\epsilon\over2}}
e^{{{i\pi\epsilon}\over4}}L^{\epsilon}(2\pi)^{-n}. \eqno(20) $$
Each integral in eq.(20) contains an `ambiguous' term of the
type eq.(13), but in the sum they cancel leaving
$$ W_1^b=-g^2C_R(2\pi)^{1-n}\Gamma(-\epsilon)L^{\epsilon}
e^{{{i\pi\epsilon}\over2}}{{2\pi^{1-{\epsilon\over2}}}\over
{\Gamma(1-{\epsilon\over2})}}\{-{2\over{\epsilon}}
2^{{\epsilon\over2}}+{\epsilon\over2}2^{{\epsilon\over2}}
[{1\over2}\ln ^22-{{\pi^2}\over{12}}]+{5\over{24}}\pi^2
\epsilon +\ln 2\}. \eqno(21)  $$
$ W_1^c $ in eq.(18) is evaluated in the same way as eq.(5),
and it gives
$$ W_1^c=g^2C_R(2\pi)^{1-n}\Gamma(-\epsilon)\cos {{\epsilon
\pi}\over2}(2L)^{\epsilon}{2\over{\epsilon}}
{{\Gamma({{1+\epsilon}\over2})}\over{\Gamma({1\over2})}}
2\pi^{1-{\epsilon\over2}}. \eqno(22) $$

\beginsection 4. The Complete Contribution from the Triangle

The complete contribution for the triangle Wilson loop to
order $ g^2 $ is the sum of the self energy graph and the
vertex graph. It is
$$  W=W_1+W_2  \eqno(23)  $$
$$  W=-g^2C_R(2\pi)^{1-n}L^{\epsilon}\{{4\over{\epsilon^2}}
+{{2i\pi}\over{\epsilon}}-{1\over{\epsilon(1-\epsilon)}}
+{{2C}\over{\epsilon}}+i\pi C-{1\over2}C +{1\over2}C^2
-{7\over{12}}\pi^2\}2\pi^{1-{\epsilon\over2}}.
\eqno(24) $$
The Euler's constant $ C $ is the relic of the expansion
of Gamma functions in powers of $ \epsilon $. Eq.(24)
agrees with the result in the Feynman gauge.

\vskip 2cm
A.A. wishes to thank Prof.J.C.Taylor for the invaluable
advise.

\beginsection  References

[1] S.\ Mandelstam, Nucl.\ Phys.\ {\bf B213} (1983) 149

\noindent
[2] G.\ Leibbrandt, Phys.\ Rev.\ {\bf D29} (1984) 1699

\noindent
[3] A.\ Andra\v si and J.\ C.\ Taylor, Nucl.\ Phys.\ {\bf B310} (1988) 222

\noindent
[4] O.\ Piguet and K.\ Sibold, Nucl.\ Phys.\ {\bf B253} (1985) 517

\noindent
[5] A.\ Andra\v si and J.\ C.\ Taylor, Nucl.\ Phys.\ {\bf B302} (1988) 123

\noindent
[6] A.\ Bassetto, I.\ A.\ Korchemskaya, G.\ P.\ Korchemsky and G.\ Nardelli,
    Nucl.\ Phys.\ {\bf B408} (1993) 62

\beginsection Figure Captions

Fig.\ 1. The self energy diagram with the gluon exchanged on
the spacelike Wilson line.

\noindent
Fig.\ 2. The vertex graph $ W_1 $. The gluon propagates between
the Mandelstam lightlike vector $ n^* $ and the spacelike
Wilson line.

\end